\documentstyle[11pt]{article}

\newcommand{\nc}{\newcommand}
\nc{\be}{\begin{equation}}
\nc{\ee}{\end{equation}}
\nc{\bea}{\begin{eqnarray}}
\nc{\eea}{\end{eqnarray}}
\nc{\RR} {\rangle }
\nc{\LL} {\langle }
\nc{\half}{\mbox{\small$\frac12$}}
\nc{\beff}{\beta_{\rm eff}}
\nc{\EW}[1]{\langle #1 \rangle}
\nc{\XX}{\phantom{0}}
\nc{\mc}{\multicolumn}

\begin{document}

\begin{titlepage}
\begin{flushright}
HUB-EP-96/12 \\
MS-TPI-96-8\\
\end{flushright}
\vspace{3ex}
\begin{center}

{\LARGE \bf
\centerline{Computing the Roughening Transition}
\centerline{of Ising and Solid-On-Solid Models}
\centerline{by BCSOS Model Matching}
}

\vspace{1.2 cm}
{\large M. Hasenbusch,}

\vspace{0.3 cm}
{\it Fachbereich Physik, Humboldt Universit\"at zu Berlin,} \\
{\it Invalidenstr.\ 110, 10099 Berlin, Germany} \\
email: hasenbus@birke.physik.hu-berlin.de

\vspace{0.8 cm}
{\large K. Pinn,}

\vspace{0.3 cm}
{\it  Institut f\"ur Theoretische Physik I, Universit\"at M\"unster,} \\
{\it  Wilhelm-Klemm-Str.\ 9, D-48149 M\"unster, Germany} \\
email: pinn@uni-muenster.de

\vspace{1. cm}
\end{center}
\setcounter{page}{0}
\thispagestyle{empty}
\begin{abstract} \normalsize
We study the roughening transition of the dual of the 2D XY model, of
the Discrete Gaussian model, of the  Absolute Value Solid-On-Solid model
and of the interface in an Ising model on a 3D simple cubic lattice. The
investigation relies on a  renormalization group finite size scaling
method  that was proposed and successfully tested a few years ago.  The
basic idea is to match the  renormalization group flow of the interface
observables  with that of the exactly solvable  BCSOS model.  Our
estimates  for the  critical couplings are $\beta_R^{XY}=1.1199(1)$,
$K_R^{DG}=0.6653(2)$ and $K_R^{ASOS}=0.80608(2)$ for the $XY$-model,
the Discrete Gaussian model and the Absolute Value Solid-On-Solid model,
respectively. For the inverse roughening temperature of the Ising
interface we find $K_R^{Ising}= 0.40758(1)$.
To the best of our knowledge, these are the most precise estimates
for these parameters published so far.
\end{abstract}
\nopagebreak
\vspace{1ex}
\end{titlepage}
\newpage

\section{Introduction}

Among the phase transitions that occur in 2D or
effectively 2D statistical systems, those of the  so called
Kosterlitz-Thouless (KT) type~\cite{KT} belong to the most challenging.
The KT phase
transition is  of infinite order: The free energy and all its
derivatives stay finite at the transition point.  Despite the relatively
simple arguments that  suggest the existence of such a transition in  a
variety of systems, a rigorous proof of the
KT nature of the phase transition in many
physically interesting systems is still lacking.
Also most of the numerical studies (many of them
Monte Carlo studies of the 2D XY model), could not provide
an unambigious confirmation of the KT
scenario. A number of references will be given in
section~\ref{SECcompa}.

The reason for the problem is the appearance of corrections to scaling
that vanish only logarithmically with the system size. Most of the
investigations  based on simulations of KT models on finite lattices
suffer from these corrections.

Note, however, that there exists at least one 2D lattice model,
which has been {\em proven} to undergo a KT transition by
exact solution. This is the Body Centered Solid-On-Solid
(BCSOS) model~\cite{beijeren77}, the configurations of which are,
up to boundary conditions, in one-to-one correspondence to those of
a special 6-vertex model~\cite{lieb,wu,baxter}, the F-model.

A few years ago we proposed a method that allows to investigate  the KT
transition of a given model by comparing its block spin renormalization
group (RG) flow (on finite lattices) with that of the
BCSOS model~\cite{old_match,thesis}.
A matching of the two RG flows at long distance
(large blocks) demonstrates that the models belong to the
same universality class, which is the KT class here.

In contrast to the usual approaches, the method introduces systematic
errors that  decay like $L^{-2}$, where $L$ is the size of the lattices
involved in the computations.

Our approach has been successfully applied to the Absolute Value
Solid-On-Solid (ASOS) model, the Discrete Gaussian (DG) model
and the dual of the
standard $XY$ model in two dimensions~\cite{old_match}. Successful
applications  to the interface of the Ising model were performed
in~\cite{thesis},  and, recently, in~\cite{IsingKL}.

In the present paper we improve on the results of ref.~\cite{old_match}
by  using larger lattice sizes and  increasing the statistics by a
factor of about 100. This became affordable both by  the availability of
faster computers and the use of  more efficient program code.

This paper is organized as follows:  In section~2 we define the models
and state the exact results for the BCSOS
model relevant to our study. We briefly discuss
the KT flow equations. Section~3 is devoted to a description of the
matching method. In section~4 we present and discuss our numerical results. A
comparison with previous estimates of the critical couplings
and non-universal parameters is presented in section~5.
Conclusions and an outlook  follow.

In the preprint version of this paper, part of the tables
is presented in a section after the Bibliography.

\section{Ising Model Interfaces and Solid-On-Solid Models}

\subsection{Ising Model Interfaces}
\label{SECmod1}

We consider the 3D Ising model on
the simple cubic lattice, with Hamiltonian
\be\label{isiham}
H = - \sum_{<x,y>} s_x s_y  \, , \qquad \quad
s_x = \pm 1 \, .
\ee
The sites of the lattice are labelled by integer coordinates
$x=(x_1,x_2,x_3)$.
The sum in eq.~(\ref{isiham})
is over all (unordered) nearest neighbour pairs of sites
in the lattice. The partition function is
\be
Z = \sum_{\{s\}} \, \exp \left( - K^I H \right) \, .
\ee
Here, the summation is over all possible configurations of
the Ising spins.
The pair interaction is normalized such that $K^I=1/(k_B T)$, where
$k_B$ denotes Boltzmann's constant,
and $T$ is the temperature.

At a critical coupling $K^I_c$ (the estimate of a recent
study~\cite{blote} is $K^I_c=0.221 6546(10)$) the infinite volume
limit of the model undergoes a second order phase transition.
For $K^I > K^I_c$, the system shows  spontaneous
breaking of the reflection symmetry.

In order to study interfaces separating extended domains of different
magnetization, we consider lattices with extension  $L$ in the $x_1$-
and $x_2$-directions and with extension $D$  in the  $x_3$-direction. We
generalize eq.~(\ref{isiham}) to
\be
  H= - \sum_{<x,y>} k_{xy} \, s_x s_y  \, .
\ee
The lattice becomes a torus by regarding the opposite boundary planes
as neighbour planes.
For the Ising spins $s$ we will apply
antiperiodic boundary conditions in $x_3$-direction, by letting
$k_{xy}=-1$ for
the  links that connect the uppermost
with the lowermost plane. For the other links we set $k_{xy}=1$.

For sufficiently large $K^I$ and large enough $L$, the imposure of
antiperiodic boundary conditions forces the  system to develop exactly
one interface, a region where  the magnetization rapidly changes sign.
This interface is parallel to a (001) lattice plane.

The Ising (001) interface undergoes a {\em roughening transition} at an
inverse temperature $K^I_R = 1 / (k_{\rm B} T_R) $ that is nearly
twice as large as the bulk transition coupling $K_c^I$
given above.\footnote{A pioneering work on this issue is
ref.~\cite{weeks73a}.}
In this work, we shall determine a new estimate for $K_R^{I}$,
and also for other parameters of the roughening transition.
For a collection of previous estimates, see section~\ref{SECcompa}.

At the roughening transition,  the large scale interface behavior
changes  from being rigid or smooth at low temperature to being rough at
high  temperature. The transition shows up in a characteristic behaviour
of various quantities. For example, in the smooth phase, the
interfacial width  stays finite when $L$ tends to infinity, while it
diverges  logarithmically with the system size in the rough
phase~\cite{kner83a,tension}. For general introductions to roughening,
see~\cite{abraham-domb,beijeren-nolden,forgacs}. For comparisons of
real life experiments with theory see, e.g., refs.~\cite{experiments}.

\subsection{Solid-On-Solid Models}
\label{SECmod2}

A fairly good approximation of the Ising interface is given  by
Solid-On-Solid (SOS) models to be introduced in this section. The SOS
approximation amounts to ignoring overhangs of the Ising interface and
bubbles in the two phases separated by the  interface.
For a review of exact results on SOS type of models, see,
e.g.~\cite{abraham-domb}. By duality~\cite{savit} and other exact
transformations (see, e.g., \cite{equiv}), SOS models have been shown to be
equivalent to a variety of other statistical models.

All SOS models that we shall consider  have in common that they are
2D lattice spin models.

Our first example of an SOS model is the Absolute-Value-Solid-On-Solid
(ASOS) model. It can be considered as the SOS approximation of an (001)
lattice plane interface  of an Ising model on a simple cubic lattice.
The model is defined by the Hamiltonian
\be
H_{ASOS} = K^{ASOS} \, |h_x - h_y| \, .
\ee
The spin variables $h_x$ take integer values. Here and in the
following, the Boltzmannian will always be $\exp(-H)$.  A factor $1/(k_B T)$,
where $k_B$ denotes Boltzmann's constant and $T$ the temperature,
is absorbed in the definition of the Hamiltonian.

We interprete the $h_x$ as heights with respect to a
certain base. For finite positive $K^{ASOS}$ the Hamiltonian will favour
that neighbouring spins take similar values. When $K^{ASOS}$ is large
enough, the surface will not fluctuate too wildly (smooth phase).
On the other hand, if $K^{ASOS}$ is below a certain critical value, the
surface becomes ``rough'', and, e.g., the surface thickness diverges
when the system size goes to infinity.

Let us now turn to the Discrete Gaussian (DG)
model. The Hamiltonian is
\be
H_{DG} = K^{DG} \, (h_x - h_y)^2 \, .
\ee
The spin variables $h_x$ take integer values. Note that the Hamiltonian
looks exactly like that of a continuous Gaussian model.
However, the
restriction of the $h_x$ to integer values introduces a nontrivial
interaction.
The Discrete Gaussian model is dual to the XY model with Villain
action~\cite{savit}.\footnote{What is called Hamiltonian
in the language of Statistical Mechanics is called action in
the framework of Euclidean Quantum Field Theory.}
This model is defined by the partition function
\be
Z_{V} = \int_{-\pi}^{\pi} \prod_x d \Theta_x \,
\prod_{< x,y >} B(\Theta_x-\Theta_y) \, ,
\ee
with
\be
B(\Theta)= \sum_{p=-\infty}^{\infty} \,
\exp \left( - \half \beta_V (\Theta - 2\pi p)^2 \right)
\ee
and
\be
\frac{1}{2\beta_V} = K^{DG} \, .
\ee
The index ``$V$'' here refers to ``Villain''.

The XY model with ``standard (cosine) action'' has the partition function
\be
Z_{XY} = \int_{-\pi}^{\pi} \prod_x d \Theta_x
\exp \left( \beta^{XY} \sum_{<x,y>} \cos(\Theta_x-\Theta_y) \right) \, .
\ee
The standard action is the mostly discussed action for an XY model.
The dual of this model is given by the partition function
\be
Z_{XY}^{SOS} = \sum_{\{h\}} \prod_{<x,y>} I_{|h_x - h_y|}(\beta^{XY}) \, ,
\ee
where the $I_n$ are modified Bessel functions. Again $h_x$ is integer.

We finally introduce the
Body Centered Solid-On-Solid (BCSOS) or F-model.
The BCSOS model
was introduced by van~Beijeren~\cite{beijeren77} as an SOS
approximation of an interface in an Ising model on a body centered cubic
lattice on a (001) lattice plane. For a detailed analysis of this model
with respect to roughening and surface structure,
see~\cite{beijeren-nolden,nolden-thesis,forgacs}.
The effective 2D lattice
splits in two sublattices like a checker board. In the original
formulation, on one of the sublattices the spins take integer values,
whereas the spins on the other sublattice take half-integer values. We
adopt a different convention: spins on ``odd'' lattice sites take values
of the form $2n+\half$, and spins on ``even'' sites are of the form
$2n-\half$, $n$ integer.
The Hamiltonian of the BCSOS model can be expressed as
\be
H_{BCSOS} =
K^{BCSOS} \sum_{[x,y]} |h_x - h_y| \, .
\ee
The sum is over next-to-nearest neighbour pairs $[x,y]$, and
nearest neighbour spins $h_x$ and $h_y$ obey the constraint
$|h_x - h_y| = 1$. Van~Beijeren~\cite{beijeren77} showed
that the BCSOS model can be transformed into
the F-model, which is a special six vertex model.
The F-model can be solved exactly
with transfer matrix methods~\cite{lieb,wu,baxter}.
The roughening transition occurs at
\be
K_R^{BCSOS} = \half \ln2 \, .
\ee
For $K \searrow K_R$, the correlation length behaves like
\be\label{xibcsos}
 \xi^{BCSOS} \simeq \frac14 \exp \left( \frac{\pi^2}{8 \sqrt{\frac12 \ln 2}}
 \, \kappa^{-\frac12} \right) \, , \quad
 \mbox{\small $\kappa = \frac{K-K_R}{K_R}$} \, .
\ee

\subsection{Renormalization Group Flow of Interface Models}

It is believed (though not proven rigorously) that,  in the vicinity of
the fixed point relevant for the KT transition,  the RG flow of SOS
models and also of the 3D Ising model interface is well described by
two parameters $\beta$ and $z$ \cite{KT}. The two parameters  are the
inverse temperature and a fugacity $z$.

The 2D Sine Gordon model is especially suited to
discuss the flow of these parameters with the length scale, since
this model contains $\beta$ and $z$ as bare parameters in its Hamiltonian:
\be
 H^{SG} =
    \frac 1 {2 \beta} \sum_{<x,y>} (\phi_x - \phi_y)^2
         - z \sum_x \cos (2 \pi \phi_x) \, ,
\ee
where the $\phi_x$ are real numbers.
For the continuum version of the model, with a momentum cutoff, one can
derive the parameter flow under infinitesimal RG transformations
\cite{KT}. It is given by
\be\label{kteq}
 \dot{{\rm x}}  = - {\rm z}^2 \; , \quad  \dot{{\rm z}} = - {\rm x}{\rm z} \;,
\ee
where  ${\rm z}  = const \cdot z$  and ${\rm x}  = \pi \beta - 2$.
$const$ depends on the particular cutoff scheme used.
The derivative is taken with respect to the logarithm of the
cutoff scale.

For large x the fugacity z flows towards ${\rm z}=0$.
The large distance behaviour of
the model is therefore that
of a Gaussian model (without a mass term). For small x,  z grows with
increasing length scale.
The theory is therefore massive, i.e., has finite correlation length.
The critical
trajectory separates these two regions in the coupling constant space.
It ends at a
Gaussian fixed point characterized by $x=0$ or $\beta=\frac{2}{\pi}$. On
the critical trajectory the fugacity vanishes as
\be
\label{fug}
 {\rm z}(t) = \frac{1}{{\rm z}_o^{-1} + t} \; ,
\ee
where $t$ is the logarithm of the cutoff scale.
Eqs.~(\ref{kteq}) are the basis for KT theory. Its immediate consequences
are derived in statistical mechanics text books, see e.g.~\cite{Itzykson}.
E.g., the correlation length in the smooth phase of
an SOS model should diverge like
\be\label{xisos}
\xi \simeq A \, \exp\left(
C \kappa^{-1/2} \right) \, , \quad
\mbox{\small $\kappa =  \frac{K-K_R}{K_R}$} \, ,
\ee
when $K \rightarrow K_R$.
We would like to emphasize another important consequence of the
KT equations that becomes apparent from the solution eq.~(\ref{fug}):
At criticality, the fugacity, which parametrizes the deviation
of the theory from a massless Gaussian model, decays with
increasing scale $t=\ln L$,
only like $(\ln L)^{-1}$. In lattice studies, $L$ is more or less
the lattice extension. Therefore, any method that is based on
an observation of the Gaussian behaviour at long
distance, suffers strongly from finite fugacity corrections
even on very large lattices.

\section{The Matching Method}
\label{uni}

The method of ref.~\cite{old_match} is closely related  to the finite
size scaling  methods proposed by Nightingale~\cite{nightingale} and
Binder~\cite{binder_fss}. No attempt is made to compute the RG flow of the
couplings explicitly, but rather the RG flow is monitored by evaluating
quantities  that are primarily  sensitive to the lowest frequency
fluctuations on a finite lattice.  One should stress that the method
does not use any of the quantitative results of KT theory. Merely
the qualitative result that there  are two important coupling parameters
in the  flow is used.

In order to separate the low frequency modes of the field
a block spin transformation~\cite{kadanoff,wilson} is used.
Blocked systems of size $l \times l$ are considered. The size $B$ of a
block (measured in units of the original lattice spacing)
is then given by
$B = L/l$, where $L$ is the linear size of the original lattice.
The linear blocking procedure defined by
\begin{equation}
\phi_X = B^{-2} \sum_{x \in X} h_x  \;,
\end{equation}
where $X$ labels square blocks of a linear extension $B$,
is used. This linear blocking rule has the half-group property
that the successive application of two transformations  with a scale-factor
of $B$  have exactly the same effect as a single transformation  with a
scale-factor of $B^2$.

Motivated by the perturbation theory of the Sine Gordon model
two types of observables are chosen:
such that are ``sensitive'' to the flow of the kinetic
term (flow of $K$), and such that are sensitive to the
fugacity (periodic perturbation of a massless Gaussian
model). For the first type of observables
\begin{equation}
A_1 = \langle  (\phi_X-\phi_Y)^2 \rangle \;,
\end{equation}
where $X$ and $Y$ are nearest neighbours on the block lattice, and
\begin{equation}\label{a2}
A_2 = \langle  (\phi_X-\phi_Z)^2 \rangle \; ,
\end{equation}
where $X$ and $Z$ are next to nearest neighbours, are chosen.
Note that these quantities are only defined for $l > 1$.
As a monitor for the fugacity the following set
of quantities (defined for $l=1,2,4$ and 8) is taken:
\begin{eqnarray}\label{a3}
  A_3 &=& \langle \cos (1 \cdot 2\pi \phi_X) \rangle \, ,
  \nonumber \\
  A_4 &=& \langle \cos (2 \cdot 2\pi \phi_X) \rangle \, .
\end{eqnarray}

\subsection{Determination of the Roughening Coupling}

There are
two parameters which have to be adjusted in order
to match the RG flow of an SOS model or of the Ising interface
with that of the critical BCSOS model:
The  coupling $K^{S}$ of the Solid-On-Solid or Ising model
and  in addition
the ratio $b=B^{S}/B^{B}=L^{S}/L^{B}$ of the lattice sizes (and hence the
block sizes) of the SOS or Ising model and the BCSOS model.
In general a $b \neq 1$
is necessary to compensate for the
different starting points of the two models on the critical RG-trajectory
\cite{old_match}.
For the proper values of the roughening coupling $K^S_R$ and
the matching constant $b$ observables of the SOS and the
BCSOS model match like
\be\label{matcheq}
A_{i,l}^{S}(b \; B,K^{S}_R) =
A_{i,l}^{B}(B,K^{B}_R)  + O(B^{-\omega}) \;,
\ee
where $i$ labels the observable and $l$
the size of the blocked lattice.
The $O(B^{-\omega})$ corrections are due to irrelevant operators.
$\omega$ is the correction to scaling exponent.
The perturbation theory of the Sine Gordon model
suggests $\omega=2$.

In order to obtain numerical estimates for the roughening coupling $K^{S}_R$
and the matching factor $b$ for a given lattice size $L^{B}$
of the BCSOS model we require that eq.~(\ref{matcheq})
is exactly fulfilled for two block observables.

We solve the system of two equations for the two observables
$A_{i,l}$ and $A_{j,l}$ numerically by first computing the $K^S_{i,l}(b)$
and $K^S_{j,l}(b)$
that solve the single equations for a given value of $b$. The intersection
of the two curves $K^S_{i,l}(b)$ and $K^S_{j,l}(b)$ gives us then the
solution of the system of two equations.  For an illustration of this method
see figures~5 and~6 of ref.~\cite{old_match}.

In~\cite{old_match} we demonstrated, that
the corrections to scaling for the observables
$A_1$ and $A_2$ for SOS models are similar to
those in the massless continuous Gaussian model. Therefore
we considered  the  ``improved'' observable $D_1$
which is defined as follows:
\be
D_1(L) = \frac{A_1^{(0)}(\infty)}{A_1^{(0)}(L)} A_1(L) \, .
\end{equation}
$A_1^{(0)}$ is computed for the massless Gaussian model defined by
\begin{equation}
H_0 = \frac12 \sum_{<x,y>} (\psi_x-\psi_y)^2  \, .
\ee
An improved quantity $D_2$ is defined analogously.
Explicit results for  $A_1^{(0)}$
and $A_2^{(0)}$ are given in table~\ref{free}.

Obviously this modification does not affect the large $L$ behaviour
since $A_{1}^{(0)}(L)=A_{1}^{(0)}(\infty)+O(L^{-2})$.
It turns out
that the results for our largest  lattice sizes are virtually
unaffected by this kind of improvement.

\subsection{Determination of Non-Universal Constants}

The matching programme also allows to determine the non-universal constants
appearing in formulae describing the divergence of observables
near the roughening transition.
In~\cite{old_match} we showed that the two non-universal parameters
$A$ and $C$ determining the critical behaviour of the correlation
length, cf.~eq.~(\ref{xisos}),
can be determined from information of the matching procedure.
For one of the models matched with the BCSOS model,
one finds
\bea
A^{SOS} &=& b_m A^{BCSOS} \\
C^{SOS} &=& q^{-1/2} C^{BCSOS}
\eea
where the parameters $A^{BCSOS}$ and $C^{BCSOS}$ can easily be extracted
from eq.~(\ref{xibcsos}).
If
\be\label{ratioslopes}
R = \frac{\partial A^{BCSOS}_{i,l}}{\partial K^{BCSOS}} \left/
\frac{\partial A_{i,l}}{\partial K} \right. \, ,
\ee
where quantities have to be taken at the roughening couplings,
is the same for all observables,
which is the case for our data,
then $q$ is given by
\be
q = \frac{K_R^{SOS} }{K_R^{BCSOS}} \, R \, .
\ee
For a more detailed discussion see ref. \cite{old_match}.

\section{Numerical Results}

We simulated the BCSOS model at its critical coupling
$K_R^{BCSOS} = \half \ln2 $ using the loop-algorithm of Evertz, Marcu
and Lana~\cite{loop}. One has to note that periodic  boundary conditions
of the F-model do not correspond to periodic boundary conditions of the
BCSOS model. Therefore updates of loops that wind
around the lattice are forbidden.

We performed $10^7$ measurements for all lattices sizes considered.
We have chosen the number of loop-updates between
two successive measurements such that the
autocorrelation times were about 1.

In addition to the observables $A_{i,l}$ we measured the interface
thickness, the total energy $E$ and $A_{i,l} \times E$, which is needed
to compute derivatives of the observables with respect to the coupling.
In order to save disc space we accumulated 1000 measurements before writing
to the file. The statistical errors were computed by jacknifing the
$10^4$ prebinned data.
As random number generator we used a combination of three shift register
generators.

We checked the reliability of the updating program by comparing the
estimates from  $10^8$ measurements for $L=4$ with the exact results
obtained by explicitly averaging over all BCSOS configurations.  Our
data are also consistent with those of ref.~\cite{IsingKL}.
In~\cite{IsingKL} lattices of size up to $L=96$ were used,  $4 \times
10^6$ measurements were performed, and the G05CAF random number
generator of the NAGLIB was used. Note also that the computer programs
of~\cite{IsingKL}
were written independently of the programs used in the present study.

The results for the BCSOS observables are summarized  in
tables~\ref{obs_bcsos1} and~\ref{obs_bcsos2}. Our estimates for the
slopes of the observables are given in tables~\ref{slope_bcsos1}
and~\ref{slope_bcsos2}. With slope we here mean the derivative of the
observables  with respect to the coupling $K^{BCSOS}$, taken at the
critical value.

We then performed the simulations for the ASOS, the DG and the dual of
the XY model. The simulations of the ASOS and DG model were done  using
a demon version~\cite{Creutz} of the Valleys-to-Mountains reflection
(VMR) algorithm~\cite{ouralg}. The simulation of the dual XY model was  done
using the  standard version of the VMR algorithm. In both cases we used
the G05CAF routine of the NAGLIB as random number generator.

Again we performed $10^7$ measurements and accumulated 1000 measurements
before writing to the file. In order to get estimates for the
observables in a neighbourhood of the simulation point, we
employed a second order Taylor expansion (note that the prebinning
forbids the use of a reweighting technique).
We thus computed the first and second derivatives of the observables,
\be
\frac{\mbox{d} A}{\mbox{d} K} = \langle H \rangle \langle A \rangle
                              - \langle H A \rangle \, ,
\ee
and
\be
\frac{\mbox{d}^2 A}{\mbox{d} K^2} =
  \langle H^2 A \rangle
- 2 \langle H \rangle \langle H A \rangle
+ 2 \langle H \rangle^2 \langle A \rangle
- \langle H^2 \rangle \langle A \rangle \, ,
\ee
for the ASOS and the DG model. In the XY case analoguos formulae were derived.
We carefully checked by comparing with results obtained from
simulations at shifted couplings
that the Taylor expansion of the $A_{i,l}$
to second order was sufficiently precise.

We performed the simulations at the previously best known estimates for
the  roughening couplings~\cite{old_match}, namely  $K_0^{DG} =0.6645$,
$K_0^{ASOS} =0.8061$   and $\beta_0^{XY} = 1.1197$.

In order to keep the length of this paper within reasonable bounds,  we
present the numerical results only for the $XY$ model,  see
tables~\ref{XY_fac} and~\ref{XY_coup}.  The tables for the other models
are available from the authors upon request.

Finally we performed the simulations for the Ising model.
The simulations were done  using the (VMR) algorithm~\cite{ouralg} adapted to
the Ising interface as discussed in refs.~\cite{prl,tension}. Following
the proposal of ref.~\cite{puetz} we encoded all $D$ spins with the same
$x_1$ and $x_2$ in a single word.  We found no way to execute operations
in parallel on the spins in one word. The advantage of the this particular
coding merly lies in considerably reduced memory access time.

The local update of the system was speeded up by replacing the standard
Metropolis update by a demon update implemented in multispin coding technique.
Also the elimination of the bubbles in the bulk phases was done using a
multispin coding technique.

In total these improvements led to a reduction of the CPU time required
for a given statistic
by a factor of about 4 compared with the code used in
refs.~\cite{prl,tension,IsingKL}.

We performed the simulations at the previously best known estimate
for the  roughening couplings~\cite{IsingKL}, $K_0^{I}=0.40754$.

We performed $3 \times 10^6$ to $8 \times 10^6$ measurements for
lattice sizes ranging from $32 \times 32 \times 31$  to
$192 \times 192 \times 31$.  Again we pre-binned the results  of 1000
measurements before writing to disc.

For the Ising interface we also computed the Taylor expansion of the
observables to second order.

In the case of the Ising interface we performed, in addition of the
matching  with the BCSOS model, the matching with the  ASOS model
at $K_R^{ASOS} = 0.80608$, which is our present estimate of the roughening
coupling for the ASOS model. For this purpose we performed additional
simulations of the ASOS model at $K_R^{ASOS} = 0.80608$ for the lattice
sizes $24,48,56,80,112$ and $160$.
The idea behind the matching with the ASOS model is that corrections
to scaling in the ASOS model are similar to those of the Ising interface.
Therefore it should be possible to obtain relaible estimates for the
roughening coupling from smaller lattice sizes this way than from the
matching with the BCSOS model.

The total CPU requirement for all our simulations accumulates
to nearly 4 years on typical modern workstations. The computers
used include all types ranging from Pentium PCs to
Alpha AXP-3000/400 and the  parallel computer
Hitachi SR2001 with 64 processors. For an overview of the lattice
sizes employed and the CPU resources needed for the
various models, see table~\ref{CPU_table}.

\begin{table}
\begin{center}
\begin{tabular}{|l|l|r|}
\hline
model & lattice sizes & CPU \\
\hline
BCSOS   & 16,24,32,40,48,56,64,80,96,112,128,160,192,224,256 & 200 d \\
XY      & 32,48,64,96,192                                    &  70 d \\
ASOS    & 32,48,64,96,128,192,256,384,512                    & 440 d \\
DG      & 16,24,32,48,64                                     &  15 d \\
Ising   & 32,48,64,96,128,192                                & 650 d \\
\hline
\end{tabular}
  \parbox[t]{.85\textwidth}
  {
  \caption[Lattices sizes and CPU requirements]
  {\label{CPU_table}
    The lattice sizes $L$ employed in our simulations of the
    various models, together with the CPU resources
    needed on an ``average modern workstation''.
    }
  }
\end{center}
\end{table}

We extracted all our estimates from the matching of the
two observables $D_2$ and $A_3$ (the last column in the tables).
Here the convergence seems optimal.

To obtain estimates for the roughening couplings and
the matching $b_m$ we employed the following

\vspace{0.3cm}
\noindent {\it \underline{RULE:} Start with the largest block lattice
size, i.e., $l=8$ (the statistical errors are the smallest here).
As a first estimate $E_1$ take the value for the largest lattice size
$L$ available. Then check whether the estimate is 2$\sigma$-compatible with
the results (also for $l=8$) for the next two smaller lattice sizes.
2$\sigma$-compatibility of two estimates $m_1$, $m_2$ with statistical
errors $e_1$, $e_2$ here means that
$|m_1-m_2| < 2 \, [e_1^2 + e_2^2]^{1/2}$.
Then also check the 2$\sigma$-consistency of $E_1$ with the  estimates
for the three largest available $L$-values for $l=4$  and $l=2$. If the
estimates are consistent,
take $E_1$ as the final estimate.  Otherwise restart the
whole procedure with $l=4$, i.e.\ take  as the first estimate $E_1$ the
value from the largest $L$ and $l=4$. If there is again a failure,
restart at $l=2$.}

\vspace{0.3cm}
Given that the corrections die out like $L^{-2}$ our {\it RULE}
ensures that systematic errors in the determination of the roughening
coupling are smaller than the statistical errors quoted.

\vspace{0.3cm}
We invite the careful reader to go through this procedure in the case
of the $XY$-model (tables~\ref{XY_fac} and~\ref{XY_coup}).
Our final estimates for the critical couplings and the matching $b_m$,
together with the values of $l$ and $L$ where the decision
procedure stopped, are given in table~\ref{res_coup}.

\begin{table}
\begin{center}
\begin{tabular}{|l||l|l||l|l|}
\hline
model & $K_R$ & from $l,L$ & $b_m$ & from $l,L$ \\
\hline
XY      & 1.1199(1)   & 4,192 & 0.93(1)  & 4,192 \\
ASOS    & 0.80608(2)  & 8,512 & 2.78(3)  & 4,512 \\
DG      & 0.6653(2)   & 2,64  & 0.32(1)  & 2,64  \\
Ising,a & 0.40759(2)  & 2,192 & 3.20(4)  & 2,192 \\
Ising,b & 0.40758(1)  & 8,192 & 3.21(3)  & 8,192 \\
\hline
\end{tabular}
  \parbox[t]{.85\textwidth}
  {
  \caption[Results for the roughening couplings and the matching $b_m$]
  {\label{res_coup}
    Our results for the roughening couplings and the matching $b_m$,
    together with the $l,L$ values that were used (see {\it RULE}).
    In the Ising model case, the index ``$a$'' refers to
    the matching with the BCSOS model, the index ``$b$'' refers
    to the matching with the ASOS model.}
  }
\end{center}
\end{table}

In order to determine the non-universal constants $A$ and $C$
we need estimates for the ratios of slopes $R$ defined
in eq.~(\ref{ratioslopes}). These ratios for the different observables
and for the different block/lattice sizes are presented (for the
$XY$ model as an example) in table~\ref{XY_rslope}.
{}From this table, and from the corresponding tables for the other
SOS models and the Ising model, we extracted by applying again our {\it RULE}
a final estimate for the ratio of slopes $R$.
Our estimates for the non-universal constants $A$ and $C$ are given
in table~\ref{res_AC}.

\begin{table}
\begin{center}
\begin{tabular}{|l|l|l|}
\hline
model & $A$ &  $C$  \\
\hline
XY      & 0.233(3)  & 1.776(4)  \\
ASOS    & 0.695(8)  & 1.099(4)  \\
DG      & 0.080(3)  & 2.438(6   \\
Ising,a & 0.80(1)   & 1.03(2)   \\
Ising,b & 0.80(1)   & 1.01(1)   \\
\hline
\end{tabular}
  \parbox[t]{.85\textwidth}
  {
  \caption[Results for the non-universal constants $A$ and $C$]
  {\label{res_AC}
    Our results for the non-universal constants $A$ and $C$.
    In the Ising model case, the index ``$a$'' refers to
    the matching with the BCSOS model, the index ``$b$'' refers
    to the matching with the ASOS model.}
  }
\end{center}
\end{table}

\section{Comparison with Previous Studies}
\label{SECcompa}

In this section we present a comparison
of our present results with some previous
estimates on the critical couplings $K_R$ and non-universal parameters
$A$ and $C$.

Let us start with the DG model.
See table~\ref{dgver} for two estimates from the seventies
and some more modern results that can be compared with
the present estimates. A comparison of the findings in~\cite{janke}
and~\cite{around} with the estimates of~\cite{old_match} was
presented in ref.~\cite{old_match}.
We would like to comment just at the apparent 1$\sigma$-incompatibility
of the present estimate for $K_R^{DG}$ with that of~\cite{old_match}.
A closer look at our data reveals that this is most likely a statistical
fluctuation:
Discarding the $L=48$ and $L=64$ lattices from the
analysis does not move our present
estimate towards the result in~\cite{old_match},
which was obtained with the same method and with lattices of size
up to $L=32$.

\begin{table}
\centering
\begin{tabular}{|l|l|l|l|l|}
\hline
\mc{1}{|c}{Authors}  &
\mc{1}{|c}{year}     &
\mc{1}{|c}{$K_R^{DG}$} &
\mc{1}{|c}{$A$} &
\mc{1}{|c|}{$C$}      \\
\hline
Swendsen \cite{swendsen77a} & 1977 & 0.77(6) & &  \\
\hline
Shugard et al.\ \cite{shugard78} & 1978 & 0.68   &            &          \\
\hline
Janke and Nather \cite{janke} & 1991  & 0.665(5)  &            &          \\
dito, fit1                    &  & 0.6657(3) & 0.1204(18) & 2.370(11)\\
dito, fit2                    &  & 0.6595(3) & 0.0287(7)  & 2.812(14)\\
\hline
Evertz et al.\ \cite{around}   & 1993 & 0.662(3)    &          &          \\
\hline
Hasenbusch et al.\ \cite{thesis,old_match}&1992/94&0.6645(6)&0.078(5)&2.44(3)\\
\hline
Hasenbusch and Pinn, {\bf this work} & 1996& 0.6653(2) & 0.080(3)& 2.438(6) \\
\hline
\end{tabular}
\parbox[t]{.85\textwidth}
{
\caption[Comparison of our DG results with previous estimates]
{ \label{dgver}
 Comparison of our results for the DG model with previous estimates.}
}
\end{table}

We now turn to the XY model. A table of previous estimates in
comparison with previously published results is given in
table~\ref{XYver}. We find our present estimates consistent
with our previous results in~\cite{old_match}.
Most of the results of the other authors, also the
estimates from series analysis by Campostrini et al.~\cite{campo}
are inconsistent with the present estimate.
We conclude that in all these cases the systematic errors
due to corrections to scaling are underestimated.

\begin{table}
\centering
\begin{tabular}{|l|l|l|l|l|}
\hline
\mc{1}{|c}{Authors}  &
\mc{1}{|c}{year}     &
\mc{1}{|c}{$K_R^{XY}$} &
\mc{1}{|c}{$A$} &
\mc{1}{|c|}{$C$}      \\
\hline
 Baillie and Gupta \cite{gupta1} &1991 & 1.1218    & 0.2129  & 1.7258   \\
\hline
 Baillie and Gupta \cite{gupta2} &1992 & 1.119(6) &  & \\
\hline
  Biferale \cite{biferale} & 1989 & 1.112(2)  &           & 1.74(20) \\
\hline
Hasenbusch et al.\ \cite{old_match}& 1992/94&1.1197(5)&0.223(13)& 1.78(2)  \\
\hline
Olsson \cite{olsson1} & 1994 & 1.12082(25) &  & 1.585(9) \\
\hline
Olsson \cite{olsson2} & 1995 & 1.12091(13) &  & 1.59(2) \\
\hline
Schultka and Manousakis \cite{schultka} & 1994 & 1.12082(16)& & 1.800(2) \\
\hline
Campostrini et al.\ \cite{campo} & 1996 & 1.1166(4)&      &  \\
\hline
Hasenbusch and Pinn, {\bf this work} & 1996 & 1.1199(1) & 0.233(3) & 1.776(4)
\\
\hline
\end{tabular}
\parbox[t]{.85\textwidth}
{
\caption[Comparison of our XY results with previous estimates]
{ \label{XYver}
    Comparison of our results for the XY model with previous estimates.}
}
\end{table}

In the case of the ASOS model, we only compare with
our previous estimate~\cite{old_match} and with
an estimate by Adler from a ninth-order low temperatur series.
The series estimate has a quite large error, but is consistent
with our result.

\begin{table}
\centering
\begin{tabular}{|l|l|l|l|l|}
\hline
\mc{1}{|c}{Authors}  &
\mc{1}{|c}{year}     &
\mc{1}{|c}{$K_R^{ASOS}$} &
\mc{1}{|c}{$A$} &
\mc{1}{|c|}{$C$}      \\
\hline
Shugard et al.\ \cite{shugard78} & 1978 & 0.81 &      &      \\
\hline
Adler \cite{adler} & 1987 & 0.787(24) & & \\
\hline
Hasenbusch et al.\ \cite{thesis,old_match}&1992/94&0.8061(3)&0.70(8)&1.14(2) \\
\hline
Hasenbusch and Pinn, {\bf this work} & 1996& 0.80608(2) & 0.695(8) & 1.099(4)
\\
\hline
\end{tabular}
\parbox[t]{.85\textwidth}
{
\caption[Comparison of our ASOS results with previous estimates]
{ \label{asosver}
    Comparison of our results for the ASOS model with previous estimates.}
}
\end{table}

We conclude this section with a comparison of the Ising interface
estimates.
Here we find that all the cited estimates of the roughening
couplings are consistent with each other. Note, however, the
large errors in the estimates that were obtained with
techniques other than the matching method.
The estimate of Mon et al.~\cite{mon88a} for $A$ seems to be the result of
a wrong method.

\begin{table}
\centering
\begin{tabular}{|l|l|l|l|l|}
\hline
\mc{1}{|c}{Authors}  &
\mc{1}{|c}{year}     &
\mc{1}{|c}{$K_R^{I}$} &
\mc{1}{|c}{$A$} &
\mc{1}{|c|}{$C$}      \\
\hline
Weeks et al.\ \cite{weeks73a} & 1973 & 0.39  & &    \\
\hline
B\"urkner and Stauffer \cite{kner83a} & 1983 & 0.396(22) & & \\
\hline
Adler \cite{adler} & 1987 & 0.404(12) & & \\
\hline
Mon et al.\ \cite{mon88a} & 1988 & 0.410(16) & 9.8(2.0) & 1.36(6) \\
\hline
Mon et al.\ \cite{mon90a} & 1990 & 0.409(4) &  &  \\
\hline
Hasenbusch \cite{thesis}&1992&0.4074(3) & &  \\
\hline
Hasenbusch et al.\ \cite{IsingKL} & 1996 & 0.40754(5)  & 0.74(2)  & 1.03(2) \\
\hline
Hasenbusch and Pinn, {\bf this work, a}
  & 1996 & 0.40759(2)  & 0.80(1) & 1.03(2) \\
Hasenbusch and Pinn, {\bf this work, b}
  & 1996 & 0.40758(1)  & 0.80(1) & 1.01(1) \\
\hline
\end{tabular}
\parbox[t]{.85\textwidth}
{
\caption[Comparison of our Ising results with previous estimates]
    { \label{isingver}
    Comparison of our results for the Ising model with previous estimates.
    The index ``a'' referst to matching with
    the BCSOS model, whereas the ``b'' means matching with the
    ASOS model.}
    }
\end{table}


\section{Conclusions and Outlook}
  By  increasing the statistics by a factor of about 100 and
  by also using larger lattices
  compared to ref.~\cite{old_match}, we obtained the most
  accurate estimates for the roughening couplings of the Ising interface, the
  ASOS model, the DG model and the 2D XY model published so far.
  In contrast to other methods
  systematical errors are under control. The matching procedure converges
  like $L^{-2}$ while other methods that rely on  analytic results derived from
  KT theory are plagued by corrections logarithmic in the lattice size.
  In addition to the precise  numbers for the roughening coupling and other
  non-universal constants the matching provides an unambigous
  confirmation of the
  KT nature of the phase transition of the models considered.
  It is interesting to compute the observables used for the matching method
  for the Sine Gordon model in perturbation theory.
  This will allow to rederive
  the KT flow equations from finite size scaling. Furthermore, it will provide
  quantitative information about the RG flow, in particular about the
  critical trajectory, which can be compared with the numerical results
  given in the present paper.

\section*{Acknowledgements}

A large part of the simulations have been performed on the SR2001 of
Hitachi Europe, Ltd.  The remaining
part of the simulations was done on workstations of the Institut
f\"ur Numerische und Instrumentelle Mathematik der Universit\"at M\"unster
and of the DAMTP, Cambridge University.
This work was supported by the Leverhulme Trust
under grant 16634-AOZ-R8 and by PPARC.

\newpage
\listoftables

\newpage

\begin{table}
\begin{center}
\begin{tabular}{|c|c|c|c|c|c|c|}
\hline
$L$& $A_{1,2}^{(0)}$&$A_{2,2}^{(0)}$&$A_{1,4}^{(0)}$& $A_{2,4}^{(0)}$
 &$A_{1,8}^{(0)}$& $A_{2,8}^{(0)}$ \\
   \hline
     16 & 0.1231474 & 0.1718750 & 0.2439180 & 0.3260905 & 0.3177271 &
0.4230146\\
     24 & 0.1205651 & 0.1689815 & 0.2341466 & 0.3156629 & 0.2815115 &
0.3834000\\
     32 & 0.1196545 & 0.1679687 & 0.2306619 & 0.3119780 & 0.2682336 &
0.3691650\\
     40 & 0.1192317 & 0.1675000 & 0.2290354 & 0.3102652 & 0.2619443 &
0.3624936\\
     48 & 0.1190016 & 0.1672454 & 0.2281480 & 0.3093327 & 0.2584828 &
0.3588444\\
     56 & 0.1188628 & 0.1670918 & 0.2276114 & 0.3087697 & 0.2563784 &
0.3566345\\
     64 & 0.1187726 & 0.1669922 & 0.2272625 & 0.3084039 & 0.2550051 &
0.3551961\\
     80 & 0.1186664 & 0.1668750 & 0.2268516 & 0.3079734 & 0.2533820 &
0.3535002\\
     96 & 0.1186088 & 0.1668113 & 0.2266280 & 0.3077394 & 0.2524963 &
0.3525768\\
    112 & 0.1185740 & 0.1667730 & 0.2264931 & 0.3075982 & 0.2519608 &
0.3520192\\
    128 & 0.1185514 & 0.1667480 & 0.2264055 & 0.3075066 & 0.2516126 &
0.3516570\\
    160 & 0.1185248 & 0.1667187 & 0.2263024 & 0.3073988 & 0.2512025 &
0.3512307\\
    192 & 0.1185104 & 0.1667028 & 0.2262464 & 0.3073402 & 0.2509794 &
0.3509989\\
    224 & 0.1185017 & 0.1666932 & 0.2262126 & 0.3073049 & 0.2508448 &
0.3508592\\
    256 & 0.1184960 & 0.1666870 & 0.2261907 & 0.3072820 & 0.2507573 &
0.3507684\\
    384 & 0.1184858 & 0.1666757 & 0.2261509 & 0.3072403 & 0.2505986 &
0.3506036\\
    512 & 0.1184822 & 0.1666718 & 0.2261370 & 0.3072258 & 0.2505429 &
0.3505459\\
$\infty$& 0.118478\XX
        & 0.166667\XX
        & 0.226119\XX
        & 0.307207\XX
        & 0.250471\XX
        & 0.350472\XX \\
\hline
\end{tabular}
  \parbox[t]{.85\textwidth}
  {
  \caption[Exact results for $A_1^{(0)}$ and $A_2^{(0)}$]
  {\label{free}
    Exact results for $A_1^{(0)}$ and $A_2^{(0)}$ as functions of
    the size of the fundamental lattice ($L$) and the size of the blocked
    lattice ($l$). The last row contains values extrapolated to
    $L=\infty$.}
  }
\end{center}
\end{table}

\begin{table}
\begin{center}
\begin{tabular}{|c|c|l|l|l|l|}
\hline
$L$& $l$ &  $A_1$ &  $A_2$ & $A_3$  &   $A_4$ \\
\hline
 16 & 1 &   &        &0.26511(30) &  0.06737(26) \\
 24 & 1 &   &        &0.24004(31) &  0.05437(25) \\
 32 & 1 &   &        &0.22572(32) &  0.04772(25) \\
 40 & 1 &   &        &0.21526(31) &  0.04341(24) \\
 48 & 1 &   &        &0.20778(30) &  0.03959(24) \\
 56 & 1 &   &        &0.20157(31) &  0.03704(24) \\
 64 & 1 &   &        &0.19656(33) &  0.03571(24) \\
 80 & 1 &   &        &0.18884(33) &  0.03252(25) \\
 96 & 1 &   &        &0.18209(34) &  0.03032(25) \\
112 & 1 &   &        &0.17797(35) &  0.02880(25) \\
128 & 1 &   &        &0.17399(33) &  0.02750(23) \\
160 & 1 &   &        &0.16704(37) &  0.02529(26) \\
192 & 1 &   &        &0.16341(40) &  0.02419(26) \\
224 & 1 &   &        &0.15935(41) &  0.02265(26) \\
256 & 1 &   &        &0.15616(42) &  0.02188(26) \\
\hline
 16 & 2 &0.085740(27)&0.117435(45)&0.22382(18)& 0.05597(13) \\
 24 & 2 &0.083135(26)&0.115240(43)&0.20229(18)& 0.04436(12) \\
 32 & 2 &0.082161(25)&0.114434(42)&0.18981(19)& 0.03851(12) \\
 40 & 2 &0.081661(24)&0.113976(41)&0.18079(19)& 0.03444(12) \\
 48 & 2 &0.081351(24)&0.113660(41)&0.17409(18)& 0.03174(12) \\
 56 & 2 &0.081129(24)&0.113497(41)&0.16888(19)& 0.02975(12) \\
 64 & 2 &0.080932(24)&0.113249(41)&0.16449(20)& 0.02812(12) \\
 80 & 2 &0.080719(24)&0.113064(41)&0.15767(20)& 0.02567(12) \\
 96 & 2 &0.080541(24)&0.112855(41)&0.15218(21)& 0.02401(12) \\
112 & 2 &0.080422(24)&0.112743(41)&0.14834(21)& 0.02275(12) \\
128 & 2 &0.080306(22)&0.112617(38)&0.14485(20)& 0.02145(11) \\
160 & 2 &0.080165(25)&0.112424(42)&0.13916(23)& 0.01983(12) \\
192 & 2 &0.080055(25)&0.112278(43)&0.13567(24)& 0.01848(12) \\
224 & 2 &0.080008(26)&0.112258(44)&0.13225(25)& 0.01763(12) \\
256 & 2 &0.079953(26)&0.112235(45)&0.12936(26)& 0.01699(13) \\
\hline
\end{tabular}
\parbox[t]{.85\textwidth}
  {
  \caption[BCSOS results for  the critical coupling, $l=1,2$]
   {\label{obs_bcsos1}
   Monte Carlo results for the $A_i$ obtained at the critical coupling
   of the BCSOS model. The block-observables $A_i$ are defined in the text.
   $L$ is the original lattice size, and $l$ is the size of the blocked
   system.}
  }
\end{center}
\end{table}

\begin{table}
\begin{center}
\begin{tabular}{|c|c|l|l|l|l|}
\hline
$L$& $l$ &  $A_1$ &  $A_2$ & $A_3$  &   $A_4$ \\
\hline
 16 &4& 0.176651(27)& 0.230216(40) & 0.221936(80) &  0.077948(65) \\
 24 &4& 0.165308(24)& 0.219807(37) & 0.194147(82) &  0.052961(63) \\
 32 &4& 0.161166(23)& 0.215927(35) & 0.179384(82) &  0.043583(61) \\
 40 &4& 0.159116(22)& 0.213952(35) & 0.169635(82) &  0.038175(60) \\
 48 &4& 0.157915(22)& 0.212736(34) & 0.162291(81) &  0.034469(59) \\
 56 &4& 0.157067(22)& 0.211902(34) & 0.156869(82) &  0.031988(59) \\
 64 &4& 0.156463(21)& 0.211246(34) & 0.152173(85) &  0.029793(59) \\
 80 &4& 0.155714(21)& 0.210486(34) & 0.145120(86) &  0.026922(58) \\
 96 &4& 0.155185(22)& 0.209912(34) & 0.139701(89) &  0.024779(59) \\
112 &4& 0.154811(21)& 0.209486(34) & 0.135546(90) &  0.023253(58) \\
128 &4& 0.154514(20)& 0.209164(31) & 0.131951(85) &  0.021861(53) \\
160 &4& 0.154099(22)& 0.208674(35) & 0.126456(96) &  0.019967(59) \\
192 &4& 0.153818(22)& 0.208360(35) & 0.12257(10)  &  0.018579(58) \\
224 &4& 0.153583(22)& 0.208080(35) & 0.11912(10)  &  0.017534(58) \\
256 &4& 0.153391(23)& 0.207869(36) & 0.11647(11)  &  0.016720(59) \\
\hline
 16 &8& 0.265665(22)& 0.325828(30) & 0.345213(51) &1.000000 \\
 24 &8& 0.216030(16)& 0.280733(24) & 0.227866(37) &0.105810(34) \\
 32 &8& 0.196646(14)& 0.264669(22) & 0.205656(38) &0.072017(32) \\
 40 &8& 0.188911(14)& 0.257187(21) & 0.189140(37) &0.055770(31) \\
 48 &8& 0.184346(13)& 0.252912(20) & 0.179067(36) &0.048696(30) \\
 56 &8& 0.181606(13)& 0.250139(20) & 0.171361(37) &0.043493(30) \\
 64 &8& 0.179710(13)& 0.248261(20) & 0.165313(38) &0.039956(30) \\
 80 &8& 0.177383(12)& 0.245856(20) & 0.156139(38) &0.034926(29) \\
 96 &8& 0.175951(12)& 0.244337(20) & 0.149442(40) &0.031601(30) \\
112 &8& 0.174995(12)& 0.243293(20) & 0.144178(39) &0.029200(29) \\
128 &8& 0.174309(11)& 0.242532(18) & 0.139959(37) &0.027326(27) \\
160 &8& 0.173355(12)& 0.241458(20) & 0.133345(41) &0.024585(29) \\
192 &8& 0.172733(12)& 0.240736(20) & 0.128466(43) &0.022659(29) \\
224 &8& 0.172270(12)& 0.240186(20) & 0.124527(44) &0.021187(29) \\
256 &8& 0.171898(13)& 0.239742(20) & 0.121338(46) &0.020028(29) \\
\hline
\end{tabular}
\parbox[t]{.85\textwidth}
  {
  \caption[BCSOS results for  the critical coupling, $l=4,8$]
  {\label{obs_bcsos2} Continuation of table~\ref{obs_bcsos1}.}
  }
\end{center}
\end{table}

\begin{table}
\begin{center}
\begin{tabular}{|c|c|l|l|l|l|}
\hline
$L$& $l$ &  $A_1$ &  $A_2$ & $A_3$  &   $A_4$ \\
\hline
  16 &1 &      &            &          4.9725(59) &  2.2881(60) \\
  24 &1 &      &            &          5.5222(88) &  2.2951(89) \\
  32 &1 &      &            &          5.888(12)  &  2.285(12) \\
  40 &1 &      &            &          6.216(14)  &  2.304(15) \\
  48 &1 &      &            &          6.463(17)  &  2.310(17) \\
  56 &1 &      &            &          6.659(20)  &  2.326(20) \\
  64 &1 &      &            &          6.839(23)  &  2.289(23) \\
  80 &1 &      &            &          7.131(29)  &  2.308(28) \\
  96 &1 &      &            &          7.340(34)  &  2.319(34) \\
 112 &1 &      &            &          7.526(40)  &  2.260(40) \\
 128 &1 &      &            &          7.714(42)  &  2.308(42) \\
 160 &1 &      &            &          8.079(59)  &  2.286(57) \\
 192 &1 &      &            &          8.202(71)  &  2.358(69) \\
 224 &1 &      &            &          8.478(81)  &  2.432(79) \\
 256 &1 &      &            &          8.719(95)  &  2.348(91) \\
\hline
  16 & 2 & -0.54495(72) &  -0.7764(12) &  3.9804(37) &  1.6789(32) \\
  24 & 2 & -0.56652(96) &  -0.8146(16) &  4.4235(56) &  1.6717(47) \\
  32 & 2 & -0.5876(12)  &  -0.8476(20) &  4.7222(73) &  1.6559(59) \\
  40 & 2 & -0.6055(15)  &  -0.8725(25) &  4.9708(89) &  1.6655(74) \\
  48 & 2 & -0.6208(17)  &  -0.8961(29) &  5.172(11)  &  1.6574(88) \\
  56 & 2 & -0.6331(20)  &  -0.9126(34) &  5.345(13)  &  1.659(10)  \\
  64 & 2 & -0.6469(23)  &  -0.9316(38) &  5.484(14)  &  1.642(12)  \\
  80 & 2 & -0.6644(28)  &  -0.9587(47) &  5.744(18)  &  1.651(14)  \\
  96 & 2 & -0.6806(33)  &  -0.9807(56) &  5.907(21)  &  1.620(17)  \\
 112 & 2 & -0.6822(39)  &  -0.9882(67) &  6.056(25)  &  1.659(20) \\
 128 & 2 & -0.7079(40)  &  -1.0227(68) &  6.189(26)  &  1.639(21) \\
 160 & 2 & -0.7199(55)  &  -1.0320(94) &  6.474(36)  &  1.641(29) \\
 192 & 2 & -0.7412(64)  &  -1.072(11)  &  6.571(44)  &  1.642(35) \\
 224 & 2 & -0.7613(75)  &  -1.103(13)  &  6.734(50)  &  1.655(40) \\
 256 & 2 & -0.7748(88)  &  -1.121(15)  &  6.939(58)  &  1.592(46) \\
\hline
\end{tabular}
\parbox[t]{.85\textwidth}
  {
  \caption[Slope of BCSOS observables for  the critical coupling, $l=1,2$]
   {\label{slope_bcsos1}
   Monte Carlo results for the derivatives of the $A_i$ with
   respect to the coupling, taken for the BCSOS model
   at the critical coupling.
   The block-observables $A_i$ are defined in the text.
   $L$ is the original lattice size, and $l$ is the size of the blocked
   system.}
  }
\end{center}
\end{table}

\begin{table}
\begin{center}
\begin{tabular}{|c|c|l|l|l|l|}
\hline
$L$& $l$ &  $A_1$ &  $A_2$ & $A_3$  &   $A_4$ \\
\hline
  16 & 4 & -1.02443(81) & -1.3944(12) & 2.9483(20) &  1.5854(17) \\
  24 & 4 & -1.0343(10)  & -1.4308(15) & 3.2837(29) &  1.4291(24) \\
  32 & 4 & -1.0626(12)  & -1.4790(18) & 3.5337(36) &  1.3832(30) \\
  40 & 4 & -1.0909(14)  & -1.5208(22) & 3.7292(44) &  1.3796(37) \\
  48 & 4 & -1.1165(17)  & -1.5573(26) & 3.8885(51) &  1.3631(44) \\
  56 & 4 & -1.1395(19)  & -1.5901(30) & 4.0296(61) &  1.3619(51) \\
  64 & 4 & -1.1596(21)  & -1.6176(33) & 4.1442(67) &  1.3456(58) \\
  80 & 4 & -1.1943(26)  & -1.6676(41) & 4.3474(85) &  1.3524(72) \\
  96 & 4 & -1.2182(31)  & -1.7022(48) & 4.491(10)  &  1.3459(87) \\
 112 & 4 & -1.2410(36)  & -1.7339(56) & 4.627(12)  &  1.3395(99) \\
 128 & 4 & -1.2668(37)  & -1.7680(58) & 4.744(12)  &  1.329(10) \\
 160 & 4 & -1.2935(50)  & -1.7997(78) & 4.942(17)  &  1.361(14)  \\
 192 & 4 & -1.3372(59)  & -1.8660(92) & 5.055(20)  &  1.372(17) \\
 224 & 4 & -1.3603(68)  & -1.897(11)  & 5.186(23)  & 1.352(20) \\
 256 & 4 & -1.3771(79)  & -1.921(12)  & 5.351(27)  & 1.330(23) \\
\hline
  16 & 8 & -1.27254(71) & -1.65259(97) & 2.6102(15) &  0.00000  \\
  24 & 8 & -1.09526(73) & -1.5480(11)  & 2.4655(15) & 1.4653(13) \\
  32 & 8 & -1.09968(81) & -1.5481(12)  & 2.7517(19) & 1.4710(16) \\
  40 & 8 & -1.10306(92) & -1.5727(14)  & 2.9006(22) & 1.3380(19) \\
  48 & 8 & -1.1188(10)  & -1.6001(16)  & 3.0496(26) & 1.3139(22) \\
  56 & 8 & -1.1344(12)  & -1.6265(18)  & 3.1721(30) & 1.2878(26) \\
  64 & 8 & -1.1514(13)  & -1.6537(20)  & 3.2715(34) & 1.2758(30) \\
  80 & 8 & -1.1828(16)  & -1.7011(25)  & 3.4528(41) & 1.2607(36) \\
  96 & 8 & -1.2067(18)  & -1.7356(29)  & 3.5904(50) & 1.2524(43) \\
 112 & 8 & -1.2338(21)  & -1.7768(33)  & 3.7300(58) & 1.2482(50) \\
 128 & 8 & -1.2553(21)  & -1.8070(34)  & 3.8241(59) & 1.2377(52) \\
 160 & 8 & -1.2906(29)  & -1.8574(46)  & 4.0041(81) & 1.2439(71) \\
 192 & 8 & -1.3261(34)  & -1.9101(54)  & 4.1256(99) & 1.2387(85) \\
 224 & 8 & -1.3503(39)  & -1.9428(62)  & 4.252(11)  & 1.240(10) \\
 256 & 8 & -1.3741(45)  & -1.9792(72)  & 4.382(13)  & 1.230(11) \\
\hline
\end{tabular}
\parbox[t]{.85\textwidth}
  {
  \caption[Slope of BCSOS observables for  the critical coupling, $l=4,8$]
   {\label{slope_bcsos2}
   Continuation of table~\ref{slope_bcsos1}.}
  }
\end{center}
\end{table}
\begin{table}
\begin{center}
\begin{tabular}{|c|c|l|l|l|l|}
\hline
$L$& $l$ &  $A_1$,$A_3$ &  $A_2$,$A_3$ & $D_1$,$A_3$  &    $D_2$,$A_3$ \\
   \hline
 32& 2& 0.9176(46) & 0.9381(65) & 0.9007(55) &0.9273(74) \\
 48& 2& 0.9239(62) & 0.9307(78) & 0.9148(69) & 0.9235(86) \\
 64& 2& 0.9206(80) & 0.9275(92) & 0.9148(86) & 0.9231(97) \\
 96& 2& 0.9362(97) & 0.9386(117)& 0.9335(101)& 0.9365(121) \\
128& 2& 0.9376(107)& 0.9356(118)& 0.9362(109)& 0.9345(120) \\
192& 2& 0.9275(136)& 0.9331(170)& 0.9266(138)& 0.9324(172) \\
   \hline
 32& 4& 0.8896(14) & 0.9035(17) & 0.8601(18) &0.8820(20) \\
 48& 4& 0.9080(20) & 0.9162(23) & 0.8923(24) &0.9045(26) \\
 64& 4& 0.9172(27) & 0.9231(31) & 0.9082(29) &0.9165(33) \\
 96& 4& 0.9263(36) & 0.9300(38) & 0.9217(38) &0.9266(40) \\
128& 4& 0.9364(40) & 0.9363(43) & 0.9340(41) &0.9345(45) \\
192& 4& 0.9287(50) & 0.9271(52) & 0.9273(51) &0.9260(52) \\
   \hline
 32& 8& 0.8421(3)  & 0.8622(4)  & 0.7731(6)  & 0.8149(6)  \\
 48& 8& 0.8760(6)  & 0.8903(7)  & 0.8329(8)  & 0.8610(9)  \\
 64& 8& 0.8890(8)  & 0.9009(10) & 0.8629(10) & 0.8825(11) \\
 96& 8& 0.9118(11) & 0.9199(13) & 0.8988(13) & 0.9109(14) \\
128& 8& 0.9185(14) & 0.9245(16) & 0.9109(16) & 0.9191(17) \\
192& 8& 0.9267(19) & 0.9297(21) & 0.9227(20) & 0.9269(22) \\
\hline
\end{tabular}
\parbox[t]{.85\textwidth}
  {
  \caption[XY results for the matching factor]
   {\label{XY_fac}
     XY results for the matching factor obtained in the
     way described after eq.~(\ref{matcheq}).
    }
  }
\end{center}
\end{table}

\begin{table}
\begin{center}
\begin{tabular}{|c|c|l|l|l|l|}
\hline
$L$& $l$ &  $A_1$,$A_3$ &  $A_2$,$A_3$ & $D_1$,$A_3$  &    $D_2$,$A_3$ \\
   \hline
  32&2&1.119982(94)&1.119545(117)&1.120352(105)&1.119773(133)\\
  48&2&1.119933(90)&1.119822(120)&1.120084(98) &1.119940(130)\\
  64&2&1.119757(95)&1.119661(114)&1.119839(100)&1.119722(119)\\
  96&2&1.119707(86)&1.119682(101)&1.119735(89) &1.119703(104)\\
 128&2&1.119831(92)&1.119852(102)&1.119845(94) &1.119863(104)\\
 192&2&1.119789(92)&1.119742(114)&1.119797(93) &1.119748(115)\\
   \hline
  32&4&1.120176(48)&1.119726(53) &1.121162(52) &1.120428(59) \\
  48&4&1.120034(46)&1.119846(51) &1.120403(50) &1.120116(55) \\
  64&4&1.119868(45)&1.119754(52) &1.120044(47) &1.119883(55) \\
  96&4&1.119791(41)&1.119735(49) &1.119860(42) &1.119785(51) \\
 128&4&1.119798(41)&1.119800(48) &1.119829(42) &1.119823(48) \\
 192&4&1.119866(42)&1.119884(47) &1.119881(43) &1.119896(48) \\
   \hline
  32&8&1.120078(32)&1.118559(33) &1.125145(34) &1.122223(39) \\
  48&8&1.120069(29)&1.119442(29) &1.121941(28) &1.120741(33) \\
  64&8&1.120095(26)&1.119701(28) &1.120979(28) &1.120310(30) \\
  96&8&1.119951(26)&1.119754(26) &1.120271(26) &1.119973(27) \\
 128&8&1.119906(24)&1.119789(26) &1.120057(25) &1.119894(27) \\
 192&8&1.119888(22)&1.119841(24) &1.119949(23) &1.119885(24) \\
\hline
\end{tabular}
\parbox[t]{.85\textwidth}
  {
  \caption[XY results for the roughening coupling]
   {\label{XY_coup}
    XY results for the roughening coupling
    obtained in the way described after eq.~(\ref{matcheq}).}
  }
\end{center}
\end{table}

\begin{table}
\begin{center}
\begin{tabular}{|c|c|l|l|l|l|}
\hline
$L$ & $l$ &
\mc{1}{|c}{$D_1$}  &
\mc{1}{|c}{$D_2$}  &
\mc{1}{|c}{$A_1$} &
\mc{1}{|c|}{$A_2$} \\
\hline
 16 & 1 &           &         & -0.4167(14) & -0.4137(17) \\
 24 & 1 &           &         & -0.4206(14) & -0.4195(25) \\
 32 & 1 &           &         & -0.4249(17) & -0.4227(33) \\
 48 & 1 &           &         & -0.4252(19) & -0.4267(48) \\
 64 & 1 &           &         & -0.4255(24) & -0.4270(67) \\
 96 & 1 &           &         & -0.4257(32) & -0.4410(99) \\
128 & 1 &           &         & -0.4267(41) & -0.439(13) \\
192 & 1 &           &         & -0.4280(59) & -0.413(19) \\
   \hline
 16 &2 &-0.4271(11) & -0.4302(11)& -0.4154(13)& -0.4046(12) \\
 24 &2 &-0.4299(13) & -0.4314(13)& -0.4194(13)& -0.4112(17) \\
 32 &2 &-0.4298(15) & -0.4304(14)& -0.4236(14)& -0.4167(23) \\
 48 &2 &-0.4288(19) & -0.4285(19)& -0.4244(17)& -0.4195(34) \\
 64 &2 &-0.4281(23) & -0.4285(24)& -0.4238(20)& -0.4271(46) \\
 96 &2 &-0.4355(33) & -0.4365(33)& -0.4263(25)& -0.4326(70) \\
128 &2 &-0.4250(41) & -0.4252(42)& -0.4281(32)& -0.4140(92) \\
192 &2 &-0.4282(60) & -0.4277(62)& -0.4276(46)& -0.412(14)  \\
   \hline
 16 &4 &-0.41961(73)& -0.42505(77) & -0.4070(13) & -0.3657(12) \\
 24 &4 &-0.42592(89)& -0.42839(90) & -0.4140(13) & -0.3962(11) \\
 32 &4 &-0.42965(98)& -0.43038(95) & -0.4182(13) & -0.4061(14) \\
 48 &4 &-0.4290(12) & -0.4291(11)  & -0.4218(13) & -0.4148(21) \\
 64 &4 &-0.4294(13) & -0.4296(13)  & -0.4213(15) & -0.4187(28) \\
 96 &4 &-0.4324(17) & -0.4323(17)  & -0.4250(17) & -0.4236(43) \\
128 &4 &-0.4296(21) & -0.4290(20)  & -0.4264(21) & -0.4177(57) \\
192 &4 &-0.4295(30) & -0.4300(29)  & -0.4281(29) & -0.4050(85) \\
   \hline
 24 &8 &-0.42239(92)& -0.42300(69) &-0.4120(18)  & -0.38303(53) \\
 32 &8 &-0.42373(70)& -0.42791(81) &-0.4087(11)  & -0.3694(18) \\
 48 &8 &-0.42871(87)& -0.43010(86) &-0.4147(13)  & -0.3952(12) \\
 64 &8 &-0.42960(99)& -0.43030(95) &-0.4177(13)  & -0.4035(15) \\
 96 &8 &-0.4320(12) & -0.4328(12)  &-0.4204(15)  & -0.4159(23)  \\
128 &8 &-0.4313(14) & -0.4311(13)  &-0.4228(15)  & -0.4192(31) \\
192 &8 &-0.4311(18) & -0.4307(17)  &-0.4267(19)  & -0.4175(48) \\
\hline
\end{tabular}
\parbox[t]{.85\textwidth}
  {
  \caption[Ratio of slopes for the XY model]
   {\label{XY_rslope}
     Monte Carlo estimates for the
     ratio of slopes for the XY and BCSOS model block observables,
     taken at the critical couplings, c.f.\ the definition in
     eq.~(\ref{ratioslopes}).}
  }
\end{center}
\end{table}

\end{document}